\documentstyle[aps,prl,multicol,amsmath,epsf]{revtex}

\begin{document}

\draft

\title{The Intrinsic Quantum Excitations of 
Low Temperature Glasses}

\author{Vassiliy Lubchenko and Peter G. Wolynes}

\address{Department of Chemistry and Biochemistry, University of California 
at San Diego, La Jolla, CA 92093-0371}

\date{\today}

\maketitle

\begin{abstract}

Several puzzling regularities concerning the low temperature
excitations of glasses are quantitatively explained by 
quantizing domain wall motions of
the random first order glass transition theory.
The density of excitations agrees with experiment and
scales with the size of a dynamically coherent region at $T_g$, 
being about 200 molecules. The phonon coupling depends on 
the Lindemann ratio for vitrification 
yielding the observed universal relation $l/\lambda \simeq 150$ between 
phonon wavelength $\lambda$ and mean free path  $l$. Multilevel behavior
is predicted to occur in the temperature range of the thermal
conductivity plateau.

\end{abstract}


\begin{multicols}{2}
\narrowtext

\newcommand{\la}{\langle}
\newcommand{\ra}{\rangle}

Decades ago, measurements of the heat capacity and thermal
conductivity of glasses at cryogenic temperatures revealed 
the presence of excitable degrees of freedom not present in perfect
crystals \cite{ZellerPohl}.  
These could be described as two 
level tunneling systems whose energies and tunneling matrix elements
were randomly distributed \cite{AHV,Phillips}. 
Coupling the tunneling systems to 
phonons explained the thermal measurements and also predicted
novel physical effects, such as the nonlinear absorption 
of sound and a phonon echo, which were later observed \cite{satur,GG}. 

The nature of the tunneling entities has remained obscure. 
Thoughtful experimentalists and theorists have noticed
puzzles  when the model quantitatively 
confronts experimental data \cite{PWA,FreemanAnderson,YuLeggett}. 
For example, the entropy 
contained in these excitations is much less than the
residual entropy frozen in at the glass transition. 
Yet, the  density of two level systems varies
only modestly from material to material. 
There is also a mysterious nearly universal relation between 
the density  of the two level systems and their coupling to 
phonons which can be 
deduced from the observation that the mean free path of 
phonons is about 150 times their wave length at low temperature
\cite{FreemanAnderson}.  
If the two level systems arise from the motions of highly localized
specific configurations of atoms, as in impurity doped crystals, 
instead of such a universal relation we would expect significant 
variation with the glass's chemical composition.  These facts lead
Yu and Leggett \cite{YuLeggett}, 
as well as others \cite{Coppersmith,BurinKagan}, to investigate the 
possibility that
the experimentally observed excitations are really highly renormalized 
collective excitations of a system of microscopic tunneling entities 
that interact strongly through the exchange of phonons. While
such a coupling seems  to be present, manifesting itself 
in spectral diffusion of the two level entities \cite{spec_diff}, 
quantitative calculations based on the interacting model suggest 
that thermodynamic manifestations 
of the interaction should be confined to ultra low temperatures 
\cite{BurinKagan,Silbey}.  This scenario then has not yet explained the 
observations which called it forth.

Here we explore an alternative view of the quantum 
excitations of a glass.  Rather than regarding the tunneling entities
as extrinsic we  quantize the excitations that are responsible
for the activated dynamical events in a liquid 
which slow as the glass transition is 
approached. These excitations are mostly frozen in at the 
liquid glass transition.  
Many aspects of these activated motions can be understood using 
the random first order transition theory of glasses \cite{PGW}.  
This theory starts from
some exactly solvable mean field glass models showing one 
step replica symmetry breaking \cite{GKS,Derrida}. The picture of activated 
motions goes beyond mean field theory by considering entropic
droplets \cite{KTW}. The theory quantitatively explains for a 
wide range of substances both the barrier heights 
\cite{XW} and  nonexponentiality of relaxations
\cite{XWbeta} observed near the glass transition.  In this picture, 
a viscous liquid or glass consists of a mosaic of frustrated domain
 walls separating regions of energetically less frustrated material.
 Each mosaic cell resembles a local minimum of the free energy,
an ``inherent structure'' if you will \cite{StillingerWeber}.
The typical size of the cells is nearly universal and does not 
vary much with the composition of the glass because it depends on 
the near universal value of the maximum vibrational amplitude 
sustainable by the glass - the Lindemann ratio. 
The mosaic length scale  depends not 
on the molecular character of the glass
but on its preparation time scale, logarithmically.

Large scale motions such as these domain wall movements have 
usually been discounted as a possible origin of the two level
systems because tunneling amplitudes decrease with
the number of independently moving entities and with the 
distance over which they move 
(see ref.\cite{Sethna,GuttmanRahman,MonAshcroft} for 
exceptions).  We will show that occasionally 
in the amorphous solid such collective tunneling can occur with 
a significant amplitude
owing to the extraordinary  multiplicity of tunneling paths 
available  for such an entity  and to the possibility
of achieving a sharp local resonance.  

The density of excitations available through collective tunneling
can be calculated using an argument based on the classical density 
of states of systems with one step broken replica symmetry 
\cite{beyond}.  
The result depends only on the glass transition 
temperature $T_g$ and the mosaic size, and explains the magnitude of
the heat capacity. The coupling of these 
tunneling excitations to phonons is set by $T_g$ 
itself, again because of the universal Lindemann
ratio.  For resonant scattering,
the material dependent factors cancel
to yield the observed
relation between phonon wavelength and mean free path.  
Our arguments show that tunneling will 
appear to be two state like over the wide-range of temperatures 
where echo experiments have been performed, but that the two state 
character breaks down at higher temperatures where 
a plateau in the thermal conductivity is observed \cite{LowTProp} 
and where single molecule experiments already give evidence for 
deviations from two-stateness \cite{BTLBO}.   

First, a simple argument for the density of excitations 
to set the stage.
The motions above the glass transition
temperature $T_g$ are rearrangements of finite sized cooperative 
regions  from one local free energy minimum to another \cite{XW}. 
The free energy cost to create a new minimum as a function of 
the droplet radius is 
$F(r) = 4 \pi \sigma(r) r^2 - \frac{4 \pi}{3} r^3 T \tilde{s}_c$.
Here, $\sigma(r) \equiv \sigma_0 a^{1/2} r^{3/2}$ is a radius
dependent surface tension whose form follows from a renormalization
group calculation and is caused by the wetting of the interface
between two particular low energy configurations by other 
possible arrangements.  
$\sigma_0$ is surface tension at the molecular length scale $a$. 
$\tilde{s}_c$ is the 
configurational entropy per unit volume, which favors creating 
other minima.
In terms of the number of particles in the droplet 
$N \equiv \frac{4 \pi}{3} (r/a)^3$, we have
$F(N)=\gamma \sqrt{N}- T s_c \, N$,
where $\gamma \equiv 2\sqrt{3 \pi} \sigma_0 a^2$, and $s_c(T)$ is 
configurational entropy per particle.
An approximate density functional calculation \cite{XW} gives 
$\gamma = \frac{3}{2} \sqrt{3 \pi} k_B T \ln(\alpha_L a^2/\pi e)$,
where $\alpha_L a^2 \sim 10^2$, the inverse square of the 
Lindemann ratio, hardly varies from substance to substance.
The maximum of the  free energy $F^\ddagger \equiv \gamma^2/4 T s_c$ is 
reached at $N_0 = (\gamma/2 T s_c)^2$ giving the typical motional
barrier. $T_g$ is set by the quenching time for the glass 
$\tau = \tau_0 e^{F^\ddagger/k_B T_g}$, where $\tau_0$ is a molecular
time scale.
At $T_g$ the system breaks up into a mosaic of regions of size $\xi$, 
where $F(N^*)=0$, giving $N^* \simeq 190$ and
$\xi \equiv {N^*}^{1/3} a$.
$\xi$ only logarithmically depends on $\tau$ and is about  
$\simeq 5.7$ molecular radii, for quenching times
of hours, independent of molecular composition.

Each mosaic cell resembles
a finite size mean field system at $T_K$ i.e. where the entropy vanishes.
The density of minima $n(\epsilon)$ 
for any system experiencing an entropy crisis is:
$n(\epsilon) d\epsilon =
\frac{d\epsilon}{k_B T_g} e^{\epsilon/k_B T_g}$.
The proportionality 
constant $1/k_B T_g$ guarantees that there is only one absolute
ground state: $\int_{-\infty}^0 d\epsilon \, 
n(\epsilon) = 1$. 
The total density of states in the sample per unit volume is
therefore 
$N(\epsilon) \simeq \frac{1}{k_B T_g \xi^3} e^{\epsilon/k_B T_g}$.
We do not distinguish between
the energy and the {\em free} energy of basins because
the zero point energy and vibrational entropy vary little from basin to basin.
At low energies, one thus gets for the conventional density of 
excitations $\bar{P} =  \frac{1}{k_B T_g \xi^3}$. For silica with 
$T_g \simeq$ 1500 K and $a \simeq$ 3.5 \AA,
this gives
$\simeq 6 \cdot 10^{45}$ J$^{-1}$ m$^{-3}$, a value typical
for many glasses \cite{LowTProp}. 
Apart from the $T_g$ variation, which has been noted 
experimentally (see e.g. \cite{RaychaudhuriPohl}), the density 
is seen to be nearly universal 
for glasses made with the same quenching rate.
By integrating the flat distribution $\bar{P}$ up to $T_g$ one finds
that the low level excitations accessible at cryogenic temperatures 
would account for less than $(a/\xi)^3 \simeq 1 \%$
of the residual entropy at $T_g$. 
\begin{figure}[t]
\epsfxsize=6cm
\centerline{\epsfbox{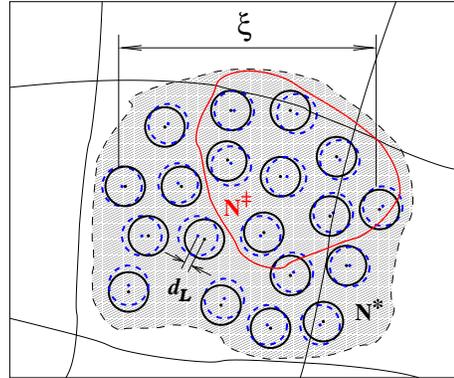}}
\caption{A schematic of a tunneling center is shown. $\xi$ is its typical  
size. $d_L$ is a typical displacement of the order of the 
Lindemann distance.
Red contour illustrates a transition state size.}
\label{fig_xi}
\end{figure}
A more complete argument demonstrates
that resonant tunneling consisting  of small
displacements of atoms within a region of size  on the order $\xi$,
as illustrated in Fig.\ref{fig_xi}, 
can indeed occur and gives the same excitation
density as the simpler argument.
We first examine the existence of resonant levels.
Consider a region from the system
quenched below $T_g$ which forms a particular local minimum.
An excitation corresponds to 
introducing a different minimum structure, encompassing $N$ molecules. 
The internal energy will be on  average
$\Delta c_p (T_g -T_K) N \simeq T_g s_c N$,  where
$\Delta c_p$ is the configurational heat capacity.
On top of the internal energy, there is a contribution
from an interface 
energy $\gamma \sqrt{N}$ due to the mismatch
with the configuration of the surrounding material. 
The distribution of excitation energies of those ``droplet''
configurations should be gaussian \cite{Derrida},
giving for the number of states higher in energy by $E$ than the
original one
$\Omega_N(E) \sim \exp \left\{s_c N - 
\frac{[E - (T_g s_c N + \gamma \sqrt{N})]^2}{2 \delta E^2 N}\right\}$. 
$\delta E^2$ is the variance in energy per particle at the glass 
transition temperature. $s_c$ is the residual entropy
that would be frozen at $T_g$ \cite{XW}. 
The entropy crisis for the bulk fixes the glass transition
temperature by $\delta E^2 = 2 s_c T_K^2$. 
The number of states at
small excitation energies grows with the excited region's size. 
Yet below a critical 
size the domain wall energy prevents resonance. The number of 
nearly resonant levels only 
becomes on the order one for $N = (\gamma/T_g s_c)^2 \Rightarrow 
N = N^*$, as before. Again,
excitation energies are distributed according to an exponential
$e^{E/k_B T_g}$, giving the same $\bar{P}$.
The multiplicity of nearly 
isoenergetic i.e. resonant configurations 
intrinsically follows from the non-equilibrium freezing
at $T_g$. 

\begin{figure}
\epsfxsize=6cm
\centerline{\epsfbox{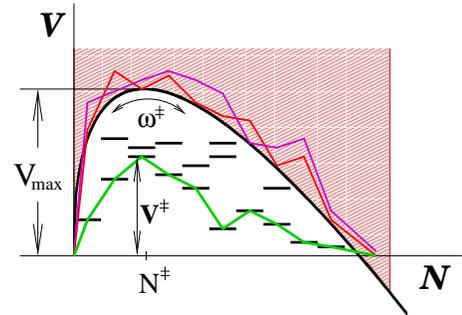}}
\caption{The black solid line shows the barrier along
the most probable path. Thick horizontal lines at low
energies and  the shaded area
at energies above the threshold
represent energy levels available at size $N$.
The red and purple line demonstrate generic paths,
green line shows the actual (lowest barrier) path, which 
whould be followed if $\hbar \omega^\ddagger < k_B T/2\pi$.}
\label{MPP}
\end{figure}

The enormous multiplicity of states at $T_g$   
not only allows resonance 
but also yields myriads of possible tunneling
paths from the initial configuration to the resonant one
with a droplet of size $N^*$ embedded in it.
Each path is a connected sequence of droplet-like 
configurations involving rearrangements of a region
containing a growing number $N < N^*$ particles.
The tunneling amplitude will be a sum over all these paths,
much like the partition sum for a random directed polymer
\cite{CookDerrida} with the weight of each path being the exponential
of its action in units of $\hbar$. 
If $\hbar$ is large, the path sum will be
dominated by a large number of paths. These would go through
a nearly continuum of paths as shown in Fig.\ref{MPP}.
This situation would correspond with a ``quantum melted glass''
\cite{SchmalianWolynes}.
On the other hand for small $\hbar$, the smallest action path only 
will contribute giving a tunneling element $e^{-S_{min}/\hbar}$.
This action varies from one resonant pair to another
and will be exponentially  distributed over a range scaling with
$k_B T_g$, giving a power law distribution for the tunneling
amplitude. 
Finding the precise statistics of the lowest action
path from the direct polymer analogy is quite complex 
since the tunneling distribution depends on 
detailed (correlated) statistics of the energy variations all
along the tunneling paths. We obtain a  sensible approximation 
by recognizing that the action will be crudely 
proportional to the largest barrier encountered
as the virtual tunneling droplet grows, as would be true
for an inverted oscillator potential.
To find the statistics of this barrier 
at any value of $N$ we must find the distribution of 
the lowest energy droplet state at $N$
(notice we neglect correlations here).
This distribution is of the same form as the one 
in $\Omega_N(E)$, except the variance is twice larger now 
because the energies of both initial and the transition states 
may fluctuate since
the system can choose where to begin to tunnel:  
$\Omega_N(V) \sim \exp \left\{s_c N - 
\frac{[V - (T_g s_c N + \gamma \sqrt{N})]^2}{4 \delta E^2 N}\right\}$.
Generally, where a resonant state exists 
the tunneling path will start by rearranging high energy local
configurations into ones with internal energies near $E_K$.
The lowest barrier likely to be encountered occurs 
when $\Omega_N(V_{mp}) \simeq 1$,
as usual in the extreme value statistics, giving
$V_{mp} = \gamma \sqrt{N} - (2\sqrt{2}-1) T_g s_c N$.
$V$ as a function of $N$ 
is shown as solid black line in Fig.\ref{MPP}. The maximum
value of $V_{mp}$ is proportional to the activation barrier 
at $T_g$, $F^\ddagger$ but smaller due to the fluctuation
of the initial energy: 
$V_{max} = F^\ddagger/(2\sqrt{2}-1) \simeq 26 T_g s_c$. 
The maximum occurs at 
$N^\ddagger \equiv N_0/(2\sqrt{2}-1)^2 \sim 14$. This is small
enough that we can expect system dependent corrections.
There will generally be states below $V_{mp}$ which
can allow tunneling as in the green line shown in Fig.\ref{MPP}. 
The distribution of barriers (and therefore actions) below $V_{max}$
follows from $\Omega_N(V)$ giving an exponential distribution
proportional to $\Omega(V^\ddagger) \sim
\exp\left\{-18 \cdot s_c + \frac{V^\ddagger}{\sqrt{2} T_g} \right\}$.
The chances to be able to tunnel to a state of precisely
the minimum resonant size is suppressed by a factor $e^{-18 s_c}$.
But to find a state which is simultaneously resonant and within
easy tunneling requires encompassing a region with only an 
$18$ additional molecules,
less than a single layer. Hence, any region of size 
$\simeq 200$ molecules will have a nearly resonant tunneling state 
within range $T_g$.
Tunneling involves simultaneous motion of all
the atoms in the droplet, and might have  
a high effective mass or even be damped owing due
to the complex rearrangements involved. It is hard to rule out
the latter possibility, but the actual effective 
mass is low since moving a domain wall 
over a molecular distance $a$ in an (imaginary) tunneling time $\tau$
involves displacing individual atoms only a Lindemann length $d_L$.
The kinetic energy associated with this motion is 
$M_w (a/\tau)^2 = N_w m (d_L/\tau)^2$, where 
$N_w \simeq (\xi/a)^2$ is the number of
molecules in the wall and $m$ is the molecular mass.
Thus the mass of the wall $M_w$ is only $ m (\xi/a)^2 (d_L/a)^2$. 
Using $(\xi/a) \simeq 5.8$ and $(d_L/a)^2 \simeq 0.01$
gives $M_w \simeq  m/3$. Computational studies of similar 
multiparticle tunneling events \cite{Sethna,GuttmanRahman}  
are consistent with the low mass obtained here.
As we shall see below, $(d_L/a)^2 \simeq k_B T_g/\rho c_s^2$,
where $\rho$ is mass density 
and $c_s$ is the speed of sound.
This gives
$M_w \simeq (\xi/a)^2 k_B T_g/c_s^2$. It follows that
the frequency of motion at the barrier top 
$\omega^\ddagger = -\partial^2 V/\partial r^2/M_w 
\simeq 1.6 (a/\xi) \omega_D$,
expressing $V$ as function of droplet's radius
$r \equiv a (3 N/4\pi)^{1/3}$. 
Since the tunneling matrix element $\Delta$ for a path with 
barrier $V^\ddagger$ is proportional to 
$e^{-\pi V^\ddagger/\hbar \omega^\ddagger}$ we obtain
a rather flat distribution for the tunneling exponent
$\log\Delta$, as is generally used to fit experiments.

We now determine the size of the coupling of the domain 
motions to the elastic (single polarization) strain field 
${\bf \nabla} \! \phi$, which has potential energy density 
$\rho c_s^2 ({\bf \nabla} \! \phi)^2/2$.
At low temperature the phonon wavelengths are long, so 
one can describe the interaction with a standard point like
term
$\sigma_z^i \, {\bf g} {\bf \nabla} \phi$, where
$\sigma_z$ is an operator of the tunneling degree of freedom.
The tunneling entities first come into existence at a temperature
$T_A$, somewhat higher than $T_g$, 
where mechanical stability of local minima to thermal 
vibrations is achieved \cite{PGW}. The phonon energies at the
microscopic scale and their coupling to the defects
will be comparable and on the order of $T_g$, giving 
$\rho c_s^2 \la ({\bf \nabla} \! \phi)^2 \ra a^3 \simeq 
\la {\bf g \nabla} \! \phi \: \sigma \ra\simeq k_B T_g$.
This sets the coupling $g$ at the molecular 
scale  $g \simeq \sqrt{\rho c_s^2 a^3 \, k_B T_g}$. The coupling
to the extended defects is weakly dependent on 
their size. To see this, in the continuum limit, we separate
the total elastic deformation tensor $u_{ij}$ into 
contributions $\phi_{ij}$ and $\pm d_{ij}/2$ due to phonon 
displacement and the tunneling motion respectively. The difference
in energy between the defect configurations in the presence
of a (longitudinal) phonon  
is then $\rho c_s^2 \phi_{ii} \int d^3 {\bf r} \, d_{ii}$, where the 
integration covers the droplet.
The coupling is therefore proportional to
a surface integral $ \int d{\bf s} \, d({\bf r})$, where 
$d({\bf r})$ are the  tunneling displacements at the edge of 
the droplet. These are random and of order $(a/\xi) d_L$
because the {\em inelastic} displacements decrease from $\sim d_L$
in the center of the droplet to zero outside.
The integral is of order $a^2 \sqrt{{N^*}^{2/3}} (a/\xi) d_L$, 
therefore $g \simeq \rho c_s^2 a^3 d_L$. 
Using  $d_L/a \simeq  \nabla \phi$
at $T_g$, one still gets $g \simeq \sqrt{\rho c_s^2 a^3 \, k_B T_g}$.

The resonant scattering from the tunneling of mosaic cells gives 
$l_{mfp}^{-1}(\omega) = \pi \frac{\bar{P} g^2}{\rho c_s^3} \omega
\tanh\left(\frac{\hbar \omega}{2 k_B T}\right)$ as 
the inverse mean free path of a phonon with frequency $\omega$
\cite{Jackle,LowTProp}. Thus, 
$\lambda_{dB}/l_{mfp} \simeq 3 \,\bar{P} g^2/\rho c^2_s$.
Combining $\bar{P} \simeq 1/k_B T_g \xi^3$  with 
the expression for the coupling constant, one
obtains $l_{mfp}/\lambda_{dB} \simeq (\xi/a)^3 \simeq 10^2$.
This ratio depends only on $\xi/a$, independent of 
molecular composition.
It is a geometrical factor reflecting the relatively low 
concentration of cooperative regions in a supercooled liquid frozen on
quenching, an almost universal
number within the random first order glass transition theory \cite{XW}.

At high temperatures the domain wall motion will become
noticeably multilevel. Ignoring damping, at a temperature
$T' \simeq \hbar \omega^\ddagger/2\pi k_B \simeq (a/\xi) T_D/2\pi$, 
the wall motion will typically be classical. This temperature  
lies within the plateau in thermal conductivity \cite{FreemanAnderson}.
Damping, which becomes considerable also at these temperatures, should
lower this estimate, as also will the fluctuations in barrier height. 
A multilevel 
system will more effectively scatter phonons, which would cause 
the plateau. Consistent with this, single molecule studies
of spectral diffusion of dyes in polymer glasses at these 
temperatures reveal 
spectral trails that wander \cite{BTLBO}, as expected for
domain walls in crystalline materials \cite{Skinner}.

{\em Acknowledgements}: We thank A.J.Leggett, A.C.Anderson and
J.Schmalian for sharing their thoughts on this problem over many
years. This work was supported by NSF grant CHE95-. It is part
of V.L.'s Ph.D. requirement at University of Illinois. 


\end{multicols}

\end{document}